\documentclass[superscriptaddress, twocolumn, aps, pra, 10pt]{revtex4-2}

\usepackage{graphicx}
\usepackage{subcaption}

\usepackage{hyperref}

\usepackage{amsmath, amssymb, amsfonts}

\begin{document}

    \title{Effective optoelectrical entanglement and strong mechanical squeezing in a multi-modulated optoelectromechanical system}
    \author{Sampreet Kalita}
        \email{sampreet@iitg.ac.in}
        \affiliation{Indian Institute of Technology Guwahati, Guwahati-781039, India}
    \author{Saumya Shah}
        \affiliation{Indian Institute of Technology Guwahati, Guwahati-781039, India}
    \author{Amarendra K. Sarma}
        \email{aksarma@iitg.ac.in}
        \affiliation{Indian Institute of Technology Guwahati, Guwahati-781039, India}

    \begin{abstract}
        We propose effective generation of entangled and squeezed states in an optoelectromechanical system comprising of a macroscopic LC electrical circuit and an optomechanical system. 
        We obtain enhanced entanglement between optical and LC circuit modes that are coupled via a common mechanical mode in the microwave regime. 
        We achieve this enhancement by a stepwise application of modulation in the laser drive, the voltage drive and the spring constant of the moveable end-mirror. 
        The maximum amount of entanglement is observed to be primarily dependent on the voltage modulation and changes slightly with the parameters of the spring constant. 
        Alongside the generated entanglement, we also study the variation of the maximum degree of squeezing in the mechanical mode for different parameter regimes.
    \end{abstract}

    \maketitle


    \section{Introduction}
        \label{sec:intro}
        
        Exploration of entangled and squeezed states of macroscopic systems has garnered tremendous interest in recent years, owing to its significance both in fundamental physics and in technology. 
        These non-classical states are studied in various platforms, most prominently in cavity optomechanical systems \cite{PhysRevA.104.053506, PhysRevA.82.021806, PhysRevA.91.013834, PhysRevA.99.043805, PhysRevLett.98.030405, AnnPhysBerl.531.1800271, PhysRevA.104.043521}.
        Entanglement, a characteristic feature of quantum mechanics, apart from shedding light onto the rudimentary structure of physical reality \cite{PhysPhysFiz.1.195}, is the chief resource for many fundamental quantum information processing tasks \cite{SBH.PhysicsQuantumInformation.Bouwmeester, RevModPhys.81.865, CUP.QuantumInformationTheory.Wilde} as well as quantum computating algorithms \cite{CUP.QuantumComputationQuantumInformation.Nielsen}.
        To this end, nanoscale optomechanical systems relying on the radiation-pressure interaction serve as an excellent platform for the study of such a quantum mechanical phenomena \cite{TaylorFrancis.QuantumOptomechanics.Bowen, RevModPhys.86.1391}.
        Their tunability in the microscopic as well as the macroscopic length-scales has been extensively utilized for the study of continuous variable entanglement \cite{PhysRevLett.98.030405, PhysRevLett.99.250401, PhysRevA.84.042342, PhysRevA.86.042306, NewJPhys.17.103037, AVSQuantumSci.3.015901}.
        Moreover, improvement in the robustness of entanglement between the optical and mechanical modes or between the mechanical modes of two optically coupled optomechanical systems have also been reported, by the application of modulated as well as squeezed laser driving \cite{PhysRevLett.103.213603, PhysRevA.89.023843, JOptSocAmB.31.1087}.

        Another characteristic feature of optomechanical systems is the ability to integrate them with solid-state electronics.
        The radiation-pressure-induced mechanical displacement in the optomechanical half \cite{RevModPhys.86.1391} of such systems induces voltages and currents in the quantum electrical elements \cite{IntJCircuitTheoryAppl.45.897, PhysRevA.104.023509}, thus enabling transfer of information between the optical and electrical components. 
        Such optoelectromechanical (OEM) systems \cite{JPhysConfSer.264.012025, NatNanotechnol.13.11} offer sufficient scalability together with the ability to leverage entanglement for efficient communication and storage schemes \cite{PhysRevA.76.042336, PhysRevA.101.042320}.
        Very recently, such works have also been extended from the microwave to the radio-frequency regime opening up new applications in communication \cite{NewJPhys.22.063041} and sensing \cite{PhysRevA.103.033516}.

        In this work, we consider an OEM system and study the entanglement between the optical and LC circuit modes mediated via a squeezed mechanical mode upon application of modulation of the laser, voltage and spring constant parameters.
        Although similar works have been carried out in the context of individual optomechanical and electromechanical systems \cite{PhysRevA.86.013820, PhysRevA.94.053807, NatCommun.11.1589}, the novelty of our work lies in the study of such modulations in a compact OEM system, the observation of \textit{optoelectrical entanglement} in the presence of multiple modulations, and the enhancement of mechanical squeezing in the presence of laser and voltage modulations, all in the microwave regime of the mechanical and circuit frequencies.
        For each observation, we provide an analytical reasoning as well as discuss the prospects of the enhancement in technological applications.
        Our work is structured as follows. 
        In Sec. \ref{sec:model}, we introduce the model and derive the corresponding dynamics of its constituent modes, formally introducing the correlation matrix of the fluctuation quadratures, and the measures in question. 
        The results obtained are presented and discussed in Sec. \ref{sec:results} and Sec. \ref{sec:conc} summarizes the work and its applications in the development of quantum technologies.


    \section{Model and Dynamics}
        \label{sec:model}
        \begin{figure}[ht]
            \centering
            \includegraphics[width=0.48\textwidth]{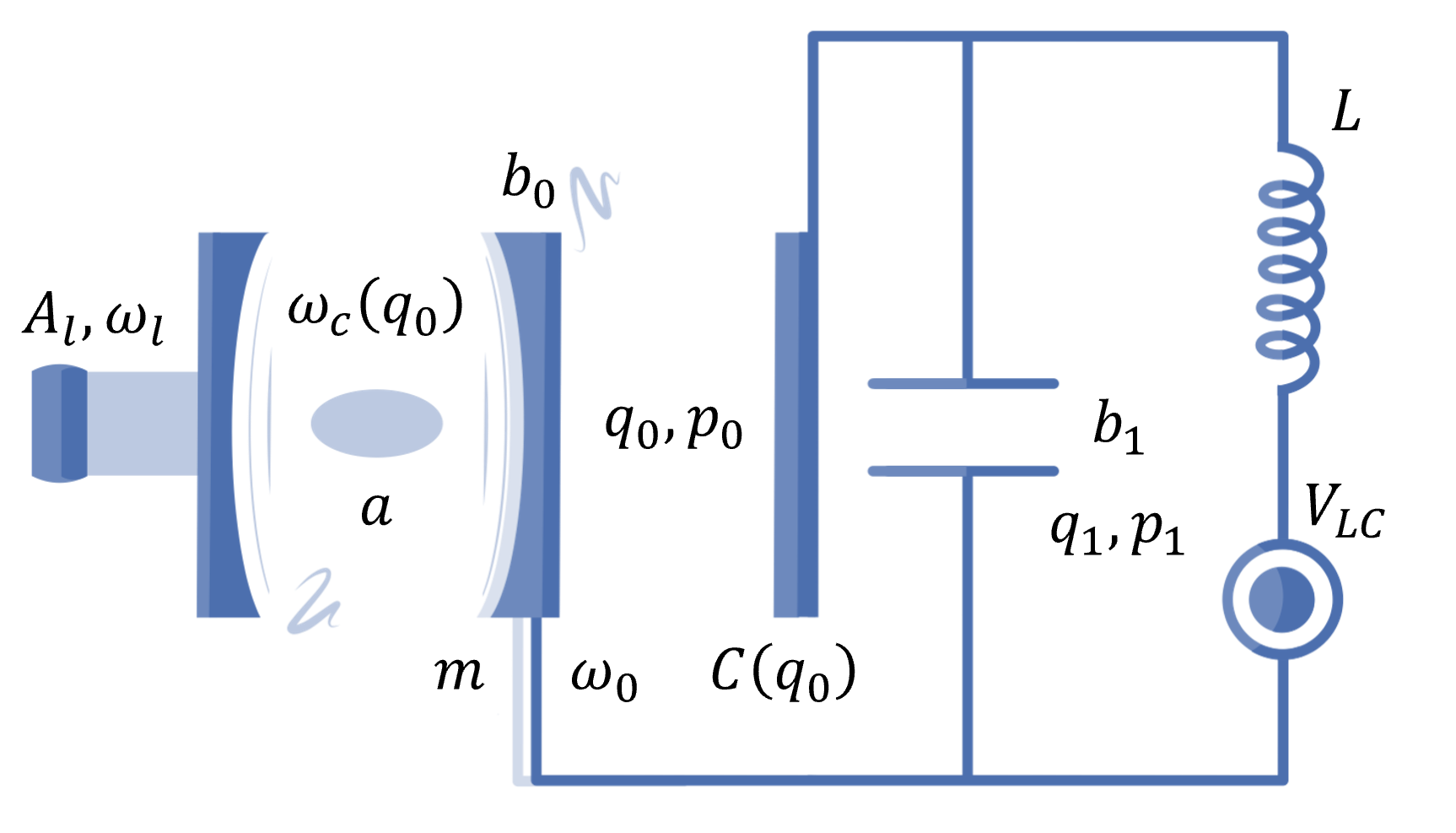}
            \caption{
                Illustration of the complete OEM system.
                An optomechanical cavity (left half) with a moveable end-mirror acts as a parallel-plate capacitor for the LC circuit (right half).
                The cavity is driven by a laser source whereas a voltage source drives the circuit.
            }
            \label{fig:model}
        \end{figure}
        
        A single optomechanical cavity is coupled to an LC circuit through its moveable end-mirror as shown in Fig. \ref{fig:model}.
        A metallic coating on the mirror serves as one plate of a parallel-plate capacitor, thereby introducing a dependency of its capacitance on the position of the mechanical mirror.
        An external voltage drives the LC circuit and amplifies the coupling between the circuit mode and the mechanical mode.
        The Hamiltonain of the complete system is given by
        \begin{align}
            \label{eqn:ham_comp}
            H = & ~ \hbar \omega_{c} a^{\dagger} a + \frac{1}{2} m \omega_{0}^{2} q_{0}^{2} + \frac{p_{0}^{2}}{2 m} + \frac{q_{1}^{2}}{2 C} + \frac{p_{1}^{2}}{2 L} \nonumber \\
            & + i \hbar A_{l} \left( a^{\dagger} e^{- i \omega_{l} t} - a e^{i \omega_{l} t} \right) - q_{1} V_{LC}.
        \end{align}
        
        Here, the first term represents the energy of the optical mode $a$ with $[a, a^{\dagger}] = 1$. 
        The second (fourth) and third (fifth) terms add up to the mechanical (LC circuit) mode energy satisfying $[q_{j}, p_{j}] = i \hbar$, with $q_{j}$ and $p_{j}$ as the position (charge) and momentum (flux) operators ($j \in \{ 0, 1 \}$ unless specified otherwise).
        The optical and mechanical are coupled \textit{dispersively}, with the cavity frequency depending on the mechanical position as $\omega_{c} ( q_{0} ) = \omega_{c} l / (l + q_{0}) \approx \omega_{c} - G q_{0}$, $l$ being the length of the cavity and $G$ the frequency-pull parameter. 
        Whereas, the electromechanical coupling is \textit{capacitive} in nature, with a resultant capacitance of $C ( q_{0} ) = C_{p} + C_{b} d / (d - q_{0})$, where $C_{p}$ and $C_{b}$ are the parasitic and base capacitances of the circuit respectively. 
        The final two terms of the Hamiltonian are laser and voltage drive energies with laser frequency $\omega_{l}$ and amplitudes $A_{l}$ and $V_{LC}$ respectively. 
        
        To analyze the collective dynamics of the cavity, mechanical and LC circuit modes, we rewrite the Hamiltonian in Eqn. \eqref{eqn:ham_comp} in terms of the corresponding dimensionless mode operators $b_{j}$'s obeying $q_{j} = q_{j_{ZP}} ( b_{j}^{\dagger} + b_{j} ) / \sqrt{2}$ and $p_{j} = i p_{j_{ZP}} ( b_{j}^{\dagger} - b_{j} ) / \sqrt{2}$, such that $[b_{j}, b_{j}^{\dagger}] = 1$.
        In the rotating frame of the laser frequency, this takes the standard form
        \begin{align}
            \label{eqn:ham_mode}
            H = & ~ \hbar \Delta^{(0)} a^{\dagger} a + \hbar \omega_{0} b_{0}^{\dagger} b_{0} - \hbar g_{0} a^{\dagger} a \left( b_{0}^{\dagger} + b_{0} \right) \nonumber \\
            & + \hbar \omega_{1} b_{1}^{\dagger} b_{1} - \hbar g_{1} \left( b_{0}^{\dagger} + b_{0} \right) \left( b_{1}^{\dagger} + b_{1} \right)^{2} \nonumber \\
            & + i \hbar A_{l} \left( a^{\dagger} - a \right) - \hbar A_{v} \left( b_{1}^{\dagger} + b_{1} \right),
        \end{align}
        where $\Delta^{(0)} = \omega_{c} - \omega_{l}$ is the cavity detuning, $\omega_{1} = 1 / \sqrt{L C ( 0 )}$ is the LC circuit resonance frequency and $A_{v} = q_{1_{ZP}} V_{LC} / ( \sqrt{2} \hbar )$ is the normalized voltage amplitude.
        The optomechanical and electromechanical coupling constants are respectively given by
        \begin{align}
            \label{eqn:coup_const}
            g_{0} = \frac{G q_{0_{ZP}}}{\sqrt{2}}, \quad g_{1} = \frac{r^{2} q_{0_{ZP}} q_{1_{ZP}}^{2}}{4 \sqrt{2} \hbar C_{b} d},
        \end{align}
        with $r = C_{b} / ( C_{p} + C_{b} )$ being the participation ratio.
        While carrying out the substitution, we have neglected the contriubtions from the terms quadratic in $q_{0} / d$.
        
        Now, from the Hamiltonain in Eqn. \eqref{eqn:ham_mode}, the quantum Langevin equations (QLEs) of the modes incorporating the fluctuation and dissipation terms can be written as
        \begin{subequations}
            \label{eqn:hle_modes}
            \begin{align}
                \dot{a} = & - i \Delta^{(0)} a - \kappa a + i g_{0} a \left( b_{0}^{\dagger} + b_{0} \right) + A_{l} \nonumber \\
                & + \sqrt{2 \kappa} a^{(in)}, \\
                \dot{b}_{0} = & - i \omega_{0} b_{0} - \gamma_{0} b_{0} + i g_{0} a^{\dagger} a + i g_{1} \left( b_{1}^{\dagger} + b_{1} \right)^{2} \nonumber \\
                & + \sqrt{2 \gamma_{0}} b_{0}^{(in)}, \\
                \dot{b}_{1} = & - i \omega_{1} b_{1} - \gamma_{1} b_{1} + 2 i g_{1} \left( b_{0}^{\dagger} + b_{0} \right) \left( b_{1}^{\dagger} + b_{1} \right) \nonumber \\
                & + i A_{v} + \sqrt{2 \gamma_{1}} b_{1}^{(in)}.
            \end{align}
        \end{subequations}
        
        Here, the input noises $\mathcal{O}^{(in)}$ are zero-mean Gaussian and delta-correlated, obeying the standard relations, $\langle {a^{(in)}}^{\dagger} ( t ) a^{(in)} ( t^{\prime} ) + a^{(in)} ( t^{\prime} ) {a^{(in)}}^{\dagger} ( t ) \rangle = \delta ( t - t^{\prime} )$ and $\langle {b_{j}^{(in)}}^{\dagger} ( t ) b_{j}^{(in)} ( t^{\prime} ) + b_{j}^{(in)} ( t^{\prime} ) {b_{j}^{(in)}}^{\dagger} ( t ) \rangle = ( 2 n_{th_{j}} + 1 ) \delta ( t - t^{\prime} )$, where $n_{th_{j}} = 1 / \{ e^{\hbar \omega_{j} / ( k_{B} T_{j} )} - 1 \}$ is the mean thermal occupancy of the mechanical and LC circuit modes at temperatures $T_{j}$.
        
        For reasonably large driving amplitudes, each mode operator can be approximated by the sum of their classical amplitudes and quantum fluctuations as $\mathcal{O} = \langle \mathcal{O} \rangle + \delta \mathcal{O}$ to obtain the classical dynamics ($\alpha = \langle a \rangle$, $\beta_{j} = \langle b_{j} \rangle$)
        \begin{subequations}
            \label{eqn:hle_class}
            \begin{align}
                \dot{\alpha} = & - \left( \kappa + i \Delta \right) \alpha + A_{l}, \\
                \dot{\beta}_{0} = & - \left( \gamma_{0} + i \omega_{0} \right) \beta_{0} + i g_{0} \alpha^{*} \alpha + i g_{1} \left( \beta_{1}^{*} + \beta_{1} \right)^{2}, \\
                \dot{\beta}_{1} = & - \left( \gamma_{1} + i \omega_{1} \right) \beta_{1} + 2 i g_{1} \left( \beta_{0}^{*} + \beta_{0} \right) \left( \beta_{1}^{*} + \beta_{1} \right) \nonumber \\
                & + i A_{v},
            \end{align}
        \end{subequations}
        and the corresponding quantum dynamics
        \begin{subequations}
            \label{eqn:hle_quant}
            \begin{align}
                \delta \dot{a} = & - \left( \kappa + i \Delta \right) \delta a + i G_{0} \left( \delta b_{0}^{\dagger} + \delta b_{0} \right) \nonumber \\
                & + \sqrt{2 \kappa} a^{(in)}, \\
                \delta \dot{b}_{0}  = & - \left( \gamma_{0} + i \omega_{0} \right) \delta b_{0} + i G_{0} \delta a^{\dagger} + i G_{0}^{*} \delta a \nonumber \\
                & + 2 i G_{11} \left( \delta b_{1}^{\dagger} + \delta b_{1} \right) + \sqrt{2 \gamma_{0}} b_{0}^{(in)}, \\
                \delta \dot{b}_{1}  = & - \left( \gamma_{1} + i \omega_{1} \right) \delta b_{1} + 2 i G_{11} \left( \delta b_{0}^{\dagger} + \delta b_{0} \right) \nonumber \\
                & + 2 i G_{10} \left( \delta b_{1}^{\dagger} + \delta b_{1} \right) + \sqrt{2 \gamma_{1}} b_{1}^{(in)}.
            \end{align}
        \end{subequations}
        
        In deriving the above set of equations, we have considered a linearized description of the system, thereby neglecting the second-order nonlinear terms in fluctuations.
        Also, four new terms, effective detuning $\Delta = \Delta_{0} - g_{0} \left( \beta_{0}^{*} + \beta_{0} \right)$, optical amplitude enhanced optomechanical coupling $G_{0} = g_{0} \alpha$, mechanical position enhanced electromechanical coupling $G_{10} = g_{1} \left( \beta_{0}^{*} + \beta_{0} \right)$ and electric charge enhanced electromechanical coupling $G_{11} = g_{1} \left( \beta_{1}^{*} + \beta_{1} \right)$, have been introduced.
        
        Finally, in terms of the dimensionless quadratures $X = ( \delta a^{\dagger} + \delta a ) / \sqrt{2}$, $Y = i ( \delta a^{\dagger} - \delta a ) / \sqrt{2}$, $Q_{j} = ( \delta b_{j}^{\dagger} + \delta b_{j} ) / \sqrt{2}$ and $P_{j} = i ( \delta b_{j}^{\dagger} - \delta b_{j} ) / \sqrt{2}$, Eqns. \eqref{eqn:hle_quant} can be written in a compact form $\dot{u} = Au + \eta$, where $u = (X, Y, Q_{0}, P_{0}, Q_{1}, P_{1})^{T}$ and $A$ is
        \begin{equation}
            \label{eqn:hle_drift}
            \left( \begin{array}{cccccc}
                - \kappa & \Delta & - 2 G_{0I} & 0 & 0 & 0 \\
                - \Delta & - \kappa & 2 G_{0R} & 0 & 0 & 0 \\
                0 & 0 & - \gamma_{0} & \omega_{0} & 0 & 0 \\
                2 G_{0R} & 2 G_{0I} & - \omega_{0} & - \gamma_{0} & 4 G_{11} & 0 \\
                0 & 0 & 0 & 0 & - \gamma_{1} & \omega_{1} \\
                0 & 0 & 4 G_{11} & 0 & - \omega_{1} + 4 G_{10} & - \gamma_{1} \\
            \end{array} \right),
        \end{equation}
        with $G_{0R}$ ($G_{0I}$) being the real (imaginary) part of $G_{0}$.
        
        Now, we emphasize on the main motivation behind deriving the above set of equations.
        The dynamics of the quantum fluctuations gives us a compact dynamical form of the correlation matrix $V_{kk^{\prime}} = \langle u_{k} u_{k^{\prime}} + u_{k^{\prime}} u_{k} \rangle / 2$ of the quadratures as $\dot{V} = A.V + V.A^{T} + D$, with $D = \mathrm{Diag} [ \kappa, \kappa, \gamma_{0} (2 n_{th_{0}} + 1), \gamma_{0} (2 n_{th_{0}} + 1), \gamma_{1} (2 n_{th_{1}} + 1), \gamma_{1} (2 n_{th_{1}} + 1)]$, from which we can analyze the two properties of interest.
        The first is the degree of squeezing of the position quadrature, which can be extracted from the corresponding diagonal element of the correlation matrix that give us the quadrature variances \cite{PhysRevA.101.053836}.
        Secondly, the reduced correlation matrices between bipartite components of the collective system facilitates the measure of entanglement between them \cite{PhysRevLett.98.030405}.
        Based on this formalism, we present and discuss the results obtained for both squeezing and entanglement as follows.

    
    \section{Results and Discussions}
        \label{sec:results}
        We begin with the analysis of an unmodulated model and successively introduce the modulations in the laser, the voltage and the spring constant.
        We verify that for the unmodulated case, neither optoelectrical entanglement nor mechanical squeezing beyond the standard quantum limit (SQL) of $0.5$ is observed.

        \begin{figure}[ht]
            \begin{subfigure}{0.23\textwidth}
                \centering
                \includegraphics[width=0.98\textwidth]{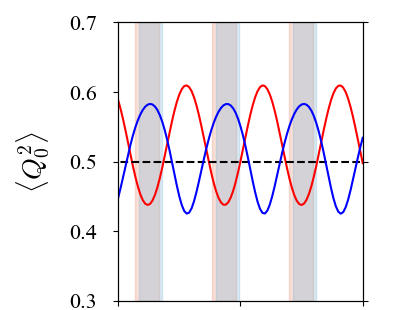}
            \end{subfigure}
            \begin{subfigure}{0.23\textwidth}
                \centering
                \includegraphics[width=0.98\textwidth]{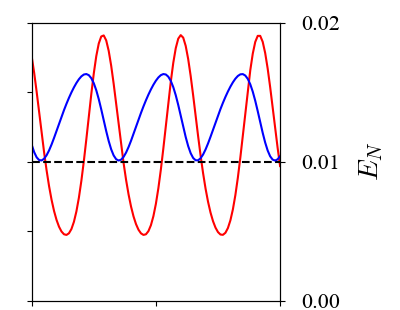}
            \end{subfigure}
            \begin{subfigure}{0.23\textwidth}
                \centering
                \includegraphics[width=0.98\textwidth]{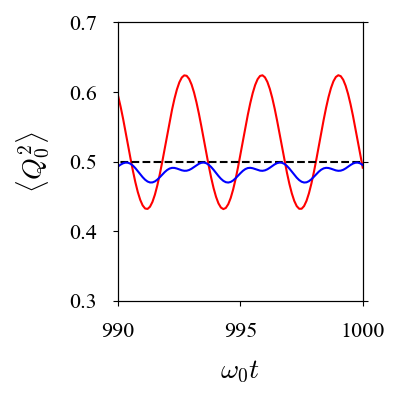}
            \end{subfigure}
            \begin{subfigure}{0.23\textwidth}
                \centering
                \includegraphics[width=0.98\textwidth]{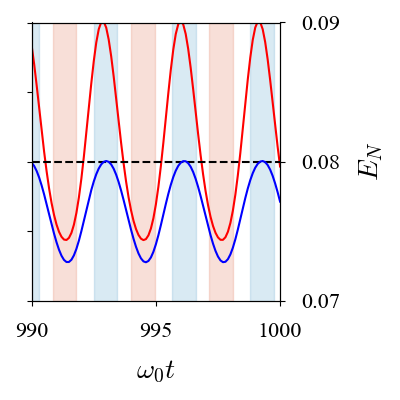}
            \end{subfigure}
            \begin{subfigure}{0.13\textwidth}
                \centering
                \includegraphics[width=0.98\textwidth]{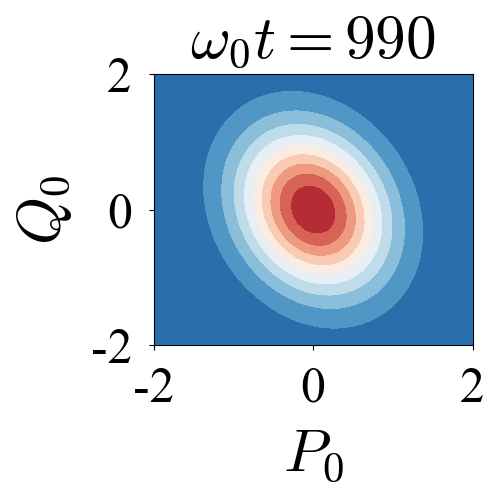}
            \end{subfigure}
            \begin{subfigure}{0.13\textwidth}
                \centering
                \includegraphics[width=0.98\textwidth]{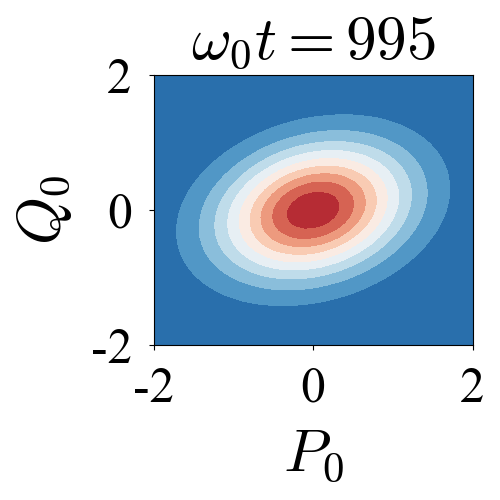}
            \end{subfigure}
            \begin{subfigure}{0.13\textwidth}
                \centering
                \includegraphics[width=0.98\textwidth]{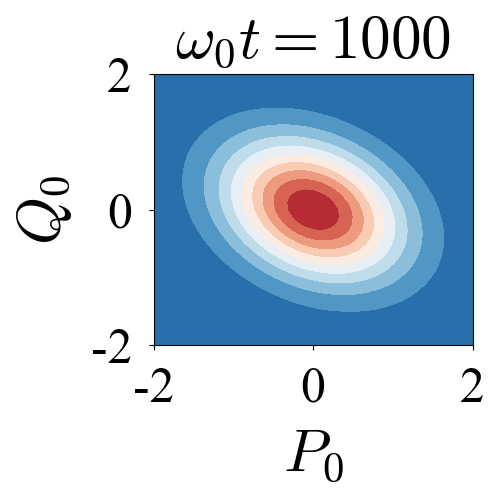}
            \end{subfigure}
            \begin{subfigure}{0.05\textwidth}
                \centering
                \includegraphics[width=0.98\textwidth]{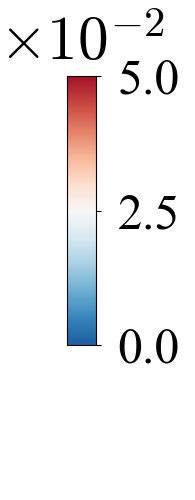}
            \end{subfigure}
            \caption{
                \textit{Line Plots:} Dynamics of mechanical squeezing (red) and optoelectrical entanglement (blue) in the presence of laser modulation.
                The bottom panels include the voltage modulation ($\Omega_{v} \neq 0$) and the right panels include the spring constant modulation ($\theta_{0} \neq 0$).
                The black dotted line represents the SQL for the mechanical position.
                The shaded regions in red and blue represent areas of high squeezing and high entanglement respectively.
                \textit{Contour Plots:} Variation of squeezing with time in the presence of all three modulations.
                The parameters used (in units of $\omega_{0}$) are $A_{l}^{(0)} = 100.0$, $A_{l}^{(\pm 1)} = 10.0$, $A_{v}^{(\pm 1)} = 50.0$, $\Delta^{(0)} = 1.0$, $\gamma_{0} = 10^{-6}$, $\gamma_{1} = 10^{-2}$, $g_{0} = 10^{-3}$, $g_{1} = 10^{-4}$, $\kappa = 0.1$, $\Omega_{l} = 2.0$, $\Omega_{v} = 2.0$, $\omega_{1} = 1.0$, $\theta_{0} = 0.5$ and $\theta_{1} = 2.0$.
            }
            \label{fig:dyna}
        \end{figure}

        \begin{figure}[ht]
            \begin{subfigure}{0.11\textwidth}
                \centering
                \includegraphics[width=0.98\textwidth]{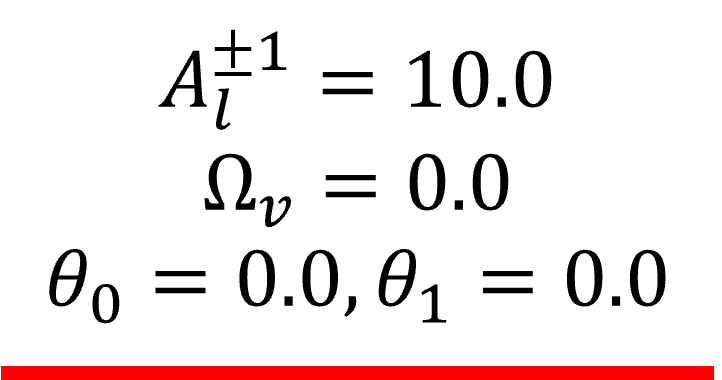}
            \end{subfigure}
            \begin{subfigure}{0.11\textwidth}
                \centering
                \includegraphics[width=0.98\textwidth]{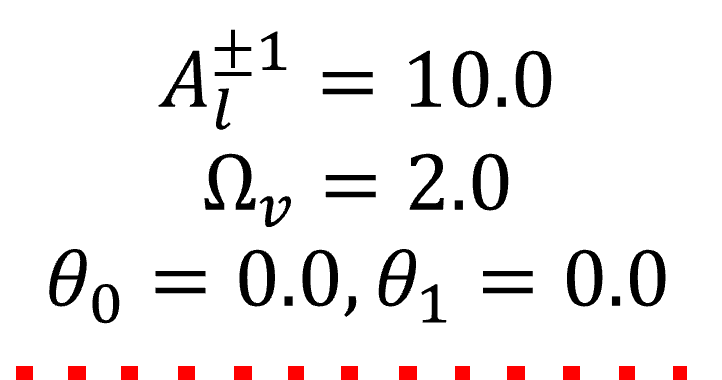}
            \end{subfigure}
            \begin{subfigure}{0.11\textwidth}
                \centering
                \includegraphics[width=0.98\textwidth]{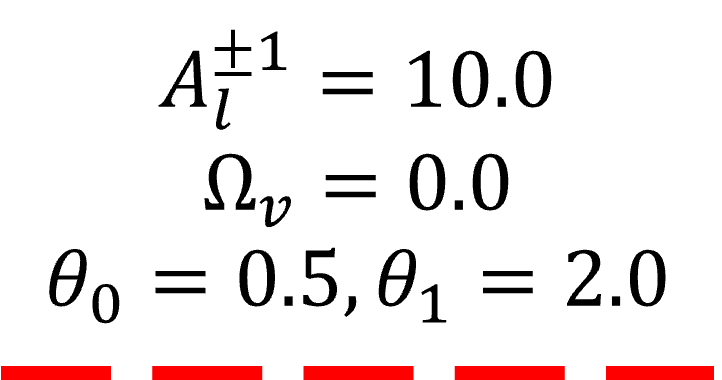}
            \end{subfigure}
            \vspace{0.4cm}
            \begin{subfigure}{0.11\textwidth}
                \centering
                \includegraphics[width=0.98\textwidth]{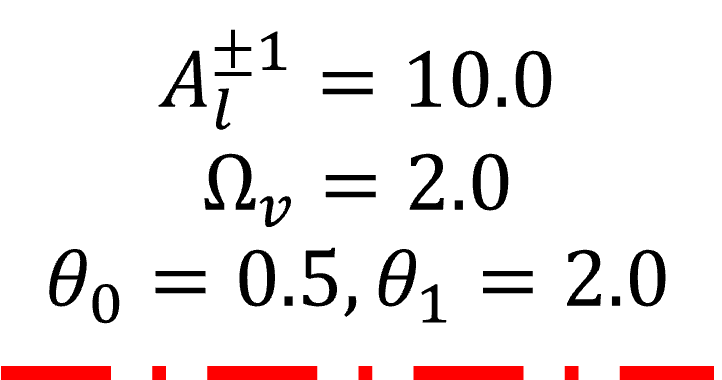}
            \end{subfigure}
            \begin{subfigure}{0.48\textwidth}
                \centering
                \includegraphics[width=0.98\textwidth]{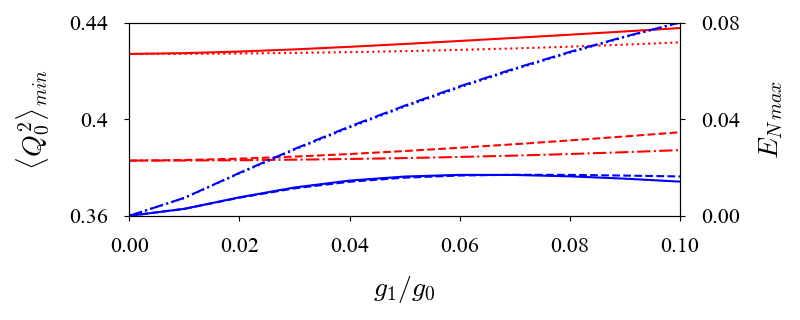}
            \end{subfigure}
            \begin{subfigure}{0.23\textwidth}
                \centering
                \includegraphics[width=0.98\textwidth]{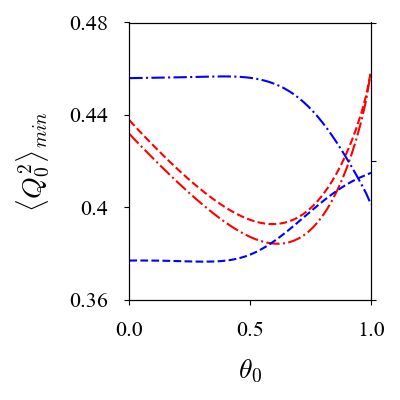}
            \end{subfigure}
            \begin{subfigure}{0.23\textwidth}
                \centering
                \includegraphics[width=0.98\textwidth]{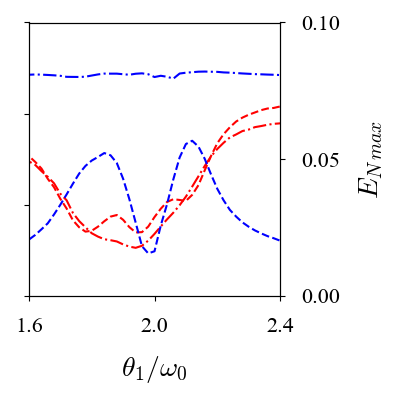}
            \end{subfigure}
            \begin{subfigure}{0.48\textwidth}
                \centering
                \includegraphics[width=0.98\textwidth]{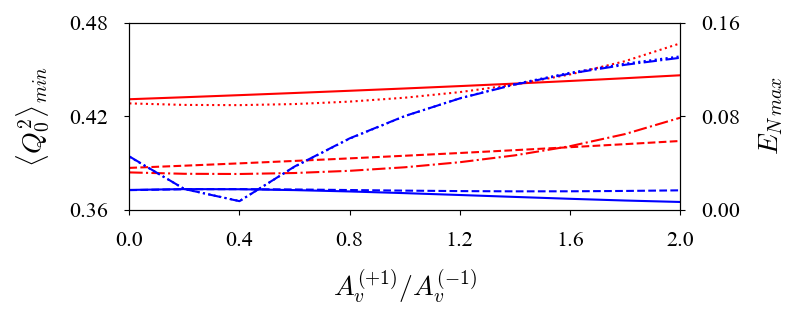}
            \end{subfigure}
            \caption{
                Variation of squeezing (red) and entanglement (blue) with change in the electromechanical coupling strength (top panel), spring contstant modulation amplitude and frequency (center left and right panels) and voltage amplitudes (bottom panel).
                Other parameters are the same as in Fig. \ref{fig:dyna}.
            }
            \label{fig:effects}
        \end{figure}

        Now, we drive the cavity by a periodically modulated laser with a fundamental sideband frequency of $\Omega_{l}$.
        As discussed in earlier works \cite{PhysRevLett.103.213603, PhysRevA.86.013820, PhysRevA.94.053807}, such a modulation not only enhances the degree of maximum entanglement achievable between the optics and mechanics, but also helps surpass the standard quantum limit of the mechanical resonator.
        To this extent, we focus on the first order sideband of the laser amplitude, which can be effectively written as $A_{l}(t) = A_{l}^{(0)} + \sum_{k = \pm 1} A_{l}^{(k)} e^{- i \Omega_{l} k t}$, as it introduces the most prominent effect.
        The modulated laser amplitude induces a periodic modulation in the amplitudes of the cavity mode, which consequently propagates from the mechanical to the LC circuit modes.
        This modulation can be visualized by looking at the steady-state behaviour of the optical mode in Eqns. \eqref{eqn:hle_class}, which reads as $\alpha_{s} = A_{l} / ( \kappa + i \Delta )$.
        The dependence of the effective parameters in Eqns. \eqref{eqn:hle_quant} on the classical amplitudes, in turn, modulates the dynamics of optoelectrical entanglement and mechanical squeezing.

        Together with the laser modulation, we consider the effect of two scenarios, first independently and then in combination as discussed below.
        First, we apply a voltage drive of the form $A_{v}(t) = \sum_{k = \pm 1} A_{v}^{(k)} e^{- i \Omega_{v} k t}$ to the circuit mode, driving it periodically at a frequency of $\Omega_{v}$.
        This generates stronger entanglement between the electrical and mechanical modes.
        Next, we gently modulate the spring constant of the mechanical mirror such that its effective frequency reads as ${\omega_{0}^{\prime}} = \omega_{0} \sqrt{\{1 + \theta_{0} \cos{(\theta_{1} t)}\}}$. 
        Fig. \ref{fig:dyna} compares the last few cycles of the dynamics of entanglement and squeezing on application of different types of modulations after a few hundred oscillation periods.
        It can be seen here that, we observe a significant amplification in the optoelectrical entanglement upon switching on the voltage modulation (bottom panels of line plots in Fig. \ref{fig:dyna}).
        Moreover, whenever modulated, squeezing of the mechanical quadrature below its standard quantum limit is observed and even higher degrees of squeezing can be achieved by modulating the spring constant of the mechanical motion at frequencies twice the mechanical frequency (right panels of line plots in Fig. \ref{fig:dyna}).
        The dynamical behaviour of the squeezing in presence of all the modulations can be visualized in the contour plots of Fig. \ref{fig:dyna}.
        In what follows, we report further effects of the electromechanical coupling and the tuneable modulation parameters on these degrees of squeezing and entanglement.

        \subsection{Effect of the Electromechanical Coupling}

            In the top panel of Fig. \ref{fig:effects}, we report the effect of the electromechanical coupling strength $g_{1}$ on the maximum values of entanglement and squeezing.
            It may be noted here that Eqn. \eqref{eqn:coup_const} introduces an upper bound to the ratio of $g_{1}$ and $g_{0}$ for experimentally feasible parameters, and thereby limits the electromechanical coupling to up to a tenth of the optomechanical one.
            None-the-less, we observe that the coupling strength plays an important role in the amount of achievable entanglement between the optical and LC circuit modes, highlighting the mediation of entanglement through the mechanical mode.
            Notably, entanglement increases substantial in the presence of voltage modulation.
            Likewise, the amount of achievable mechanical squeezing increases manifold upon application of modulation for the spring constant, and this further enhances in the presence of voltage modulation.
            For a better insight, we analyze the effect of this mechanical modulation.

        \subsection{Effect of the Mechanical Modulation}

            The effect of changes in the spring constant parameters $\theta_{0}$ and $\theta_{1}$ are reported in the central panels of Fig. \ref{fig:effects}.
            We observe that the squeezing is maximum when the frequency of modulation is twice the mechanical frequency and its modulation amplitude is close to half.
            The corresponding variations in entanglement mark enhancement in the overall achievable entanglement in the presence of all types of modulation.
            We do not go into the detailed analysis of the structure of these variations as it has already been analytically explored in the context of optomechanical systems in Ref. \cite{PhysRevA.86.013820}.
            Our primary focus lies in discussing the relation between the individual variations in squeezing and entanglement.
            In this line, a fascinating observation is the observation of alternate periodic areas of high squeezing and high entanglement in presence of all types of modulation (refer bottom right line plot of Fig. \ref{fig:dyna}) unlike the one observed when both maximum entanglement and sqeezing are in sync (refer top left line plot of Fig. \ref{fig:dyna}).
            This effect is prominent in the presence of voltage modulation, which plays a crucial role in the amplification of optoelectrical entanglement.
            We, therefore, focus on the effect of the modulating sidebands in the voltage drive in the next section.

        \subsection{Effect of Assymetric Sidebands}

            The effect of unequal sideband amplitudes $A_{v}^{(-1)}$ and $A_{v}^{(+1)}$ of the voltage drive is reported in the bottom panel of Fig. \ref{fig:effects}.
            Therefore, the optoelectrical entanglement is amplified for higher values of $A_{v}^{(+1)} / A_{v}^{(-1)}$ with only slight changes when the spring modulation is switched on.
            A similar observation in the context of optomechanical systems by considering the rotating wave approximation is provided in Ref. \cite{PhysRevA.101.053836}.
            In our system, however, the enhancement can be primarily attributed to the amplification of the sideband close to the effective detuning of the cavity, which leads to stronger interaction between the optical, mechanical and LC circuit modes, thereby amplifying the effective optoelectrical entanglement.
            Also, a unique variation in the squeezing is observed, where after a certain value of the ratio of sideband amplitudes, higher unevenness in the sidebands decrease the degree of maximum squeezing compared to the absence of modulation.
            This can be attributed to the destructive interference between the induced current arising out of the mechanical position and the tendency of charge accumulation between the plates owing to amplification in the modulated voltage.


    \section{Conclusion}
        \label{sec:conc}
        We systematically studied the effect of multiple layers of modulation on the generation of optoelectrical entanglement in a hybrid OEM system and reported the enhancement in the achievable entanglement in the presence of such modulations for experimentally feasible parameters.
        For a MHz-scale mechanical oscillator and LC circuit resonator in the microwave regime, we observed that the laser and voltage modulations greatly increased the amount of achieveable optoelectrical entanglement.
        A direct advantage of our study is transduction between optical and microwave regimes which paves the path towards the development of novel schemes for quantum radars \cite{PhysRevLett.114.080503} that rely on OEM converters for their operation.
        Another key output of our result \textemdash alternate observations of high squeezing and high entanglement regimes \textemdash leads towards alternate storage and retrieval of information in the coupled optoelectrical and optomechanical components.


    \section*{Acknowledgement}
        S. Kalita would like to acknowledge MHRD, Government of India for providing financial support for his research via the PMRF May 2020 Lateral Entry scheme.

    \bibliography{references}

\end{document}